\documentclass[a4paper,11pt]{article}
\usepackage{jheppub} 
\usepackage{amssymb} 
\usepackage{arydshln}
\usepackage{bm}

\newcommand{\beq}{\begin{equation}}
\newcommand{\beqnn}{\begin{equation*}}
\newcommand{\eeq}{\end{equation}}
\newcommand{\eeqnn}{\end{equation*}}
\newcommand{\bea}{\begin{eqnarray}}
\newcommand{\eea}{\end{eqnarray}}
\newcommand{\bal}{\begin{align}}
\newcommand{\eal}{\end{align}}
\newcommand{\chisq}{\ensuremath{\chi^2}}
\newcommand{\eq}[1]{eq.~(\ref{#1})}
\newcommand{\Eq}[1]{Eq.~(\ref{#1})}
\newcommand{\eqs}[2]{eqs.~(\ref{#1})-(\ref{#2})}

\newcommand{\half}{\tfrac{1}{2}}
\newcommand{\bi}{\begin{itemize}}
\newcommand{\ei}{\end{itemize}}

\newcommand{\mat}[9]{\begin{pmatrix}#1&#2&#3\\#4&#5&#6\\#7&#8&#9\end{pmatrix}}

\newcommand{\lam}{\lambda}
\newcommand{\zz}{\bm{z_0}}

\newcommand{\zb}{\bm{\overline{z}_0}}
\newcommand{\zql}{\lambda_q}
\newcommand{\zul}{\lambda_u}
\newcommand{\zdl}{\lambda_d}
\newcommand{\zq}{\bm{\lambda_q}}
\newcommand{\zu}{\bm{\lambda_u}}
\newcommand{\zd}{\bm{\lambda_d}}

\newcommand{\rhoz}{\rho_0}
\newcommand{\etaz}{\eta_0}
\newcommand{\sz}{s_0}
\newcommand{\cz}{c_0}

\newcommand{\pibyeight}{\tfrac{\pi}{8}}

\newcommand{\pibyfour}{\tfrac{\pi}{4}}
\newcommand{\pibytwo}{\tfrac{\pi}{2}}

\newcommand{\tpibyeight}{\tfrac{3\pi}{8}}

\newcommand{\rhob}{\overline{\rho}}
\newcommand{\etab}{\overline{\eta}}

\newcommand{\smallsim}{\smallsym{\mathrel}{\sim}}
\newcommand{\xx}{\ensuremath{\lambda_0}}
\newcommand{\xxSq}{\ensuremath{\lambda_0^2}}
\newcommand{\xxFour}{\ensuremath{\lambda_0^4}}
\newcommand{\xxFive}{\ensuremath{\lambda_0^5}}

\newcommand{\Sigz}{\ensuremath{{\zd+\zu}}}

\newcommand{\Dzsq}{{\bm \delta}{\bm \lambda^2}}

\newcommand{\ms}{m_s}
\newcommand{\mc}{m_c}
\newcommand{\mb}{m_b}
\newcommand{\mt}{m_t}
\newcommand{\betaz}{\beta_0}
\newcommand{\bz}{\betaz}
\newcommand{\bt}{\tilde{\beta}}

\newcommand{\Jcp}{\JCP}
\newcommand{\JCP}{J_{CP}}
\renewcommand{\aa}{A_0}
\newcommand{\aaSq}{A_0^2}
\newcommand{\bb}{b}

\newcommand{\dd}{c_d}
\newcommand{\du}{c_u}
\newcommand{\dq}{c_q}

\newcommand{\fc}{f_c}
\newcommand{\DD}{{\cal D}}
\newcommand{\ra}{\rightarrow}
\newcommand*{\ditto}{\texttt{"}}
\newcommand{\Mq}{M_q}
\newcommand{\Mqz}{M_q^0}
\newcommand{\flam}{f_{\lam}}
\newcommand{\frho}{f_{\rho}}

\newcommand{\fA}{f_A}

\newcommand{\ruFit}{2.00}
\newcommand{\rcFit}{3.46}
\newcommand{\rdFit}{4.97}
\newcommand{\rsFit}{1.968}
\newcommand{\lamFit}{0.22499}
\newcommand{\AFit}{0.876}
\newcommand{\rbFit}{0.1519}
\newcommand{\etabFit}{0.3477}
\newcommand{\chisqFit}{1.01}

\newcommand{\nDof}{2}
\newcommand{\ruPull}{0.00}
\newcommand{\rcPull}{-0.01}
\newcommand{\rdPull}{0.03}
\newcommand{\rsPull}{0.02}
\newcommand{\lamPull}{-0.03}
\newcommand{\APull}{-0.09}
\newcommand{\rbPull}{-0.76}
\newcommand{\etabPull}{-0.65}

\newcommand{\rcRenorm}{$3.46\pm0.03$}

\newcommand{\rsRenorm}{$1.968\pm0.008$}

\newcommand{\ARenorm}{$0.877^{+0.017}_{-0.016}$}

\newcommand{\ruExp}{$2.00\pm0.05$}
\newcommand{\rcExp}{$3.67\pm0.04$}
\newcommand{\rdExp}{$4.97\pm0.06$}
\newcommand{\rsExp}{$1.854\pm0.007$}
\newcommand{\lamExp}{$0.22501\pm0.00068$}
\newcommand{\AExp}{$0.826^{+0.016}_{-0.015}$}
\newcommand{\rbExp}{$0.1591\pm0.0094$}
\newcommand{\etabExp}{$0.3523^{+0.0073}_{-0.0071}$}
\newcommand{\alphaExp}{$91.6\pm1.4$}

\newcommand{\gammaExp}{$65.7\pm1.3$}

\makeatletter
\newcommand{\smallsym}[2]{#1{\mathpalette\make@small@sym{#2}}}
\newcommand{\make@small@sym}[2]{%
  \vcenter{\hbox{$\m@th\downgrade@style#1#2$}}%
}
\newcommand{\downgrade@style}[1]{%
  \ifx#1\displaystyle\scriptstyle\else
    \ifx#1\textstyle\scriptstyle\else
      \scriptscriptstyle
  \fi\fi
}
\makeatother

\title{Unitarity triangle angles explained: a predictive new quark mass matrix texture}

\author[a]{P.F.~Harrison}
\author[b]{W.G.~Scott}
\affiliation[a]{Department of Physics, University of Warwick,\\Coventry CV4 7AL, United Kingdom.}
\affiliation[b]{Rutherford Appleton Laboratory,\\ Chilton, Didcot OX11 0QX, United Kingdom.}
\emailAdd{p.f.harrison@warwick.ac.uk}
\emailAdd{william.scott@stfc.ac.uk}

\abstract{
We propose a novel quark mass matrix texture-pair with five free parameters, which fits the four quark mass ratios $m_s/m_b$, $m_d/m_b$, $m_c/m_t$, $m_u/m_t$, and the four 
CKM quark mixing observables. The matrices each have one texture zero, but the main innovation here is a ``geometric'' ansatz exploiting a pair of small complex expansion parameters, based on the geometry of the Unitarity Triangle. The fit to the observables is in good agreement with current experimental values renormalised to $\sim\!\!10^4$ TeV, and offers decisive tests against future high-precision measurements of the unitarity triangle angles at the weak scale. We identify two novel symmetries of these mass matrices which explain the phenomenologically-successful relations $\alpha\equiv\phi_2\simeq\pibytwo$ and $\beta\equiv\phi_1\simeq\pibyeight$.
}

\arxivnumber{2501.18508}

\begin{document}
\maketitle
\flushbottom

\section{Introduction}
There has been a long history of textures proposed for the quark mass/Yukawa matrices (MMs), both as stand-alone ansatze, and/or as the result of (or motivation for) models which can constrain them. The traditional guiding principle for such textures has been to reduce the number of free parameters (a priori 36) to a number smaller than the ten observables, in order to find predictive constraints among the measurable quantities, especially relationships between the quark masses and mixings \cite{fritzsch1,fritzsch2,fritzsch3,texZero}, which each manifest marked hierarchical structure. Methods have included using symmetries to enforce texture-zeroes \cite{zeroesSymms}, and/or mechanisms to produce a hierarchy among the matrix elements \cite{fnMech}. Recently, ideas have turned to the realm of modular forms \cite{modular0,modular1}, where again approaches have been found to enforce zeroes \cite{modularZeroes} and/or hierarchies \cite{modularHierarchies}. With the continuing advance of experimental determinations, many  ansatze, e.g.~\cite{fritzsch2,fritzsch3}, are now excluded by data, see e.g.~\cite{bb}. We introduce here a new MM texture based purely on the CKM and quark mass phenomenology, which is modestly predictive, fits the data, and has some novel and potentially interesting features.

The closeness of the (beauty-down) Unitarity Triangle (UT) \cite{UT1,UT2} to a right-angled triangle is striking \cite{masinaAndSavoy,rightUT,AKMS}, and it is also notable that its three internal angles \cite{pdg},\footnote{We use the measured value for $\beta$ and the CKM fit values \cite{pdg} for $\alpha$ and $\gamma$, since these give the smallest uncertainties.}
\beq
\alpha\equiv\phi_2=(91.6\pm1.4)^{\circ};
\quad\beta\equiv\phi_1=(22.6\pm0.4)^{\circ};
\quad\gamma\equiv\phi_3=(65.7\pm1.3)^{\circ},
\label{alphagamma}
\eeq
are very close to the aesthetically appealing triplet \cite{gershon}:
\beq
(\alpha,\beta,\gamma)\simeq(\alpha_0,\beta_0,\gamma_0)
\equiv(\pibytwo,\pibyeight,\tpibyeight).
\label{abc}
\eeq

We consider the possibility that these suggestive angles may be significant, not merely as numerical coincidences, but taken together, as a clue to what may lie behind the quark flavour structure. So, we set-out from the start to build them into a MM texture and we later show that they can be enforced by a pair of symmetries among the MMs. This approach is successful, at least in terms of reducing the number of free parameters needed to describe the data, as we will see. We first emphasise that our ansatz does not assure exact equality in  \eq{abc}, but rather, we adopt analogous equalities into the (arguably more fundamental) MMs, which yield, after diagonalisation, the approximate equalities seen at the level of the UT in the equation.

\section{The texture}
The proposed new texture is common to both the up and down quark MMs, constraining them relative to those of the SM. Without loss of generality, we take the MMs to be Hermitian. We adopt the popular texture zero in the $13$ and $31$ elements \cite{fritzsch2,fritzsch3,texZero}, so that the smallest mixing elements result from the non-commutation of the (small) diagonalising transformations in the $12$ and $23$ subspaces. In order both to introduce viable hierarchies among the MMs' elements, and to include $CP$ violation, we adopt two small, complex expansion parameters,\footnote{Throughout this paper, where ambiguity can arise, bold font is used to denote complex quantities, and the corresponding light font symbols, their moduli. Complex conjugation is indicated by the conventional $^*$ symbol.} $\zu$ and $\zd$, powers of which are inserted into the up and down quark MMs respectively, each playing a role similar to Wolfenstein's $\lambda$ parameter 
in the CKM matrix \cite{wolfenstein}, but distinct for each of the two MMs. Thus, we express 
both MMs as follows:
\begin{align}
\renewcommand*{\arraystretch}{1.2}
\Mq=n_q\Mqz\equiv n_q\mat{\dq\zql^4}{\bb\,\zq^3}{0}
			{\smallsim}{\bb\zql^2}{\aa\zq^2}
			{0}{\smallsim}{1},\quad q=u,d,
\label{tex1}
\end{align}
with the normalisations $n_q\simeq m_t, m_b$ for $q=u,d$ respectively, the real coefficients $(\aa,\bb,\du,\dd)$ are of magnitude $\lesssim {\cal O}(1)$, and the `$\smallsim$' symbol means ``the complex conjugate of the diagonally-opposite entry''. The equality in the coefficients between the 12 and 22 elements is conventional, any a priori difference having been absorbed into the normalisations of $\zq$, by virtue of the different powers involved.

The overall normalisations of the $\zq~(q=u,d)$ are free to vary, but we will show later that their complex sum has magnitude close to the value of Wolfenstein’s expansion parameter:
\beq
|\zd+\zu|\equiv\xx=\lambda+{\cal O}(\lam^3).
\label{xydef}
\eeq
The absolute phases of the $\zq$ are unobservable, specific choices corresponding to particular phase conventions related by simultaneous, common rephasings of both (see below). An inequality among the $\zq$ magnitudes, $\zul<\zdl$, is anticipated to ensure that the up-like mass hierarchy is more extreme than the down-like one, and in order to reproduce the triplet of experimental results for the UT angles, \eq{abc}, we take the ratio of the $\zq$ to be the exact complex \emph{constant}:
\beq
\frac{\zu}{\zd}=-i\,\tan{\pibyeight}.\label{zrat}
\eeq
The constraint, \eq{zrat}, is the most significant innovation in this paper. Specifically, in our texture, $(\zu/\zd)^*$ is identifiable in leading approximation with the complex ratio of the two sloped sides of the UT as we show below. Thus $\arg{(-\zu/\zd)}$ corresponds to the UT angle $\alpha$, the constrained value ensuring $\alpha\simeq\pibytwo$; the ratio $|\zu/\zd|$ then corresponds similarly to $\tan{\beta}$ (while simultaneously determining the asymmetry between the mass hierarchies in the two charge sectors). Thus \eq{zrat} applied in our texture, \eq{tex1}, yields \eq{abc}. Moreover, the precise values for the ratio asserted by \eq{zrat} imply and are implied by two novel symmetries of the $\Mqz$, as we will also show.

The MM texture, \eq{tex1}, uses five free parameters, $\aa$, $\bb$, $\du$, $\dd$ and $\xx$ to fit the three quark mixing angles, the $CP$ phase, and the four quark mass ratios, making it modestly predictive. A more general parameterisation would allow the $(\aa,\bb)$ parameters to be different for $q=u,d$, but we find that \emph{it is possible to fit all the quark mass and mixing observables with the parameter-pair $(\aa,\bb)$, common to both mass matrices}.
Thus a weak isospin (up/down) reflection symmetry is respected by the real parameters ($\aa$, $\bb$), and broken only by the $\zq$, the pair of parameters $\dq$ which govern $m_u$ and $m_d$, and by the matrix normalisations, $n_q$ in \eq{tex1}.

\section{Leading-order solution}
As usual we diagonalise the $\Mq$ by the unitary transformations:
\beq
U^q\Mq U^{q\dag}={\rm diag}(m_1^q,m_2^q,m_3^q), \quad q=u,d,
\eeq
but since the diagonalisation is insensitive to the normalsation of $M_q$, we work with the dimensionless $\Mqz$, \eq{tex1}.
The off-diagonal elements are small compared with the corresponding diagonal ones, so we can use small-angle approximations. We find at leading order in $\xx$ that the largest eigenvalue of the $\Mqz$ is unity, while the second eigenvalue is given by:
\beq
\frac{m_2^q}{m_3^q}=\bb \zql^2, \quad q=u,d,\label{secondRatio}
\eeq
also at leading order in $\xx$. Consecutive $2\times 2$ diagonalisations can be made in the 23 and 12 sub-spaces leading to the $U^q$ each being composed of a product of two 2D (generally complex) rotation matrices. Since rotations about different axes do not commute, small entries are induced in the 13 elements of $U^q$. At leading order:
\bal
U^q
	\simeq\mat{1}{\pm\zq}{\aa\zq^3}
       			{\mp\zq^*}{1}{-\aa\zq^2}
       			{0}{\aa\zq^{2*}}{1},\quad q=u,d,\label{diagMat}
\end{align}
where the upper sign\footnote{\Eq{diagMat} applies exactly as shown, only in the phase convention in which $\zd$ is real and $\zu$ is imaginary, the extra factor $-1$ in the $q=u$ case arising as the phase of two cancelled powers of $\zu$.} is for the $q=u$ case and the lower for $q=d$.
Combining both diagonalising matrices, we obtain the CKM matrix at leading order in $\xx$ (note \eq{xydef}):
\beq
V_{CKM}=U^uU^{d\dag}\simeq\mat	{1}{\Sigz}{\aa\zu\,\Dzsq}
						     	{-(\Sigz)^*}{1}{\aa\Dzsq}
       							{\aa\zd^*\,\Dzsq^*}{-\aa\Dzsq^*}{1}
		\sim   \mat	{1}{\xx}{\aa\xx^2\,\zu}
				{-\xx}{1}{\aa\xx^2}
       				{\aa\xx^2\,\zd^*}{-\aa\xx^2}{1},
\label{vckm}
\eeq
where\footnote{$|\zd-\zu|=|\zd+\zu|$ due to the $\zu/\zd$ phase difference of $\pibytwo$.}
\beq
\Dzsq\equiv {\bm \zd^2-\bm \zu^2}\Rightarrow |\Dzsq|=\xx^2.
\eeq

The first matrix in \Eq{vckm} is independent of the choice of phase convention, while the second applies in the PDG phase convention where $V_{us}$ and $V_{cb}$ are real. We can now compare the off-diagonal elements with their Wolfenstein equivalents, to obtain in this convention:
\begin{align}
\zu&\simeq \lam(\rho-i\eta)=V_{ub}/A\lam^2\label{z1u}\\
\zd&\simeq \lam(1-\rho+i\eta)=V_{td}^*/A\lam^2\label{z1d}\\
\aa&\simeq A,\label{A0A}
\end{align}
where $\lambda$, $\rho$, $\eta$ and $A$ are the Wolfenstein parameters \cite{wolfenstein} (and \eq{A0A} explains our choice of symbol $\aa$ in \eq{tex1}). This confirms that at leading order, 
$\zu^*$ and $\zd^*$ are each proportional to one of the ``sloped'' sides of the UT (since, in the PDG phase convention, $V_{ud}\simeq V_{tb}\simeq 1$). We can thus immediately justify the claims made below \eq{zrat}:
\begin{align}
\frac{\zu}{\zd}&=\frac{V_{ub}}{V_{td}^*},\quad{\rm cf.}~\alpha\equiv\arg\left(-\frac{V_{td}V_{tb}^*}{V_{ud}V_{ub}^*}\right),\\
\Rightarrow \arg\left({-\frac{\zu}{\zd}}\right)&=\arg\left({-\frac{V_{td}}{V_{ub}^*}}\right)\simeq\alpha,~~{\rm and}~~\left| \frac{\zu}{\zd} \right| \simeq \tan{\beta},\label{justification}
\end{align}
where the definition of $\alpha$ is from the PDG \cite{pdg}, and the last near-equality is a consequence of the near right-angle in the resulting UT.

\begin{figure}[t]
\begin{center}
\includegraphics[scale=1.20]{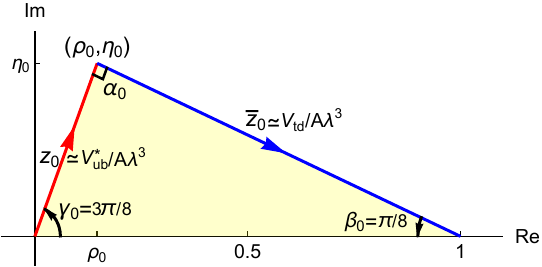}
\caption{\label{fig1}
Leading-order Unitarity Triangle used as input to textures. The UT extracted after diagonalisation of the mass matrices is congruent to it at leading order in small quantities. The angles $(\alpha_0,\bz,\gamma_0)$ are defined in \eq{abc}, and $(\rhoz,\etaz)=(0.146,0.354)$.}
\end{center}
\end{figure}

Still in the PDG phase convention where $\zd+\zu$ is real (and equals $\xx$), it is useful to define the complex constants:\footnote{The ``bar'' in $\zb$ is simply to identify this variable as the one-complement of $\zz$ and should not be confused with the bar appearing above the ``reduced'' Wolfenstein parameters $\rhob$ and $\etab$.}
\bal
\zz&\equiv\zu^*/\xx= i\sz\, e^{-i\beta_0}=\,\rhoz+i\etaz,\label{zAns0}\\
\zb&\equiv\zd^*/\xx=\cz\, e^{-i\beta_0}=\,1-\zz,
\label{zpluszb}
\end{align}
where $\bz\equiv\pibyeight$ was defined in \eq{abc}, and:\footnote{$s_0=\half\sqrt{2-\sqrt{2}}=0.383$ and $c_0=\half\sqrt{2+\sqrt{2}}=0.924$.}
\beq
\sz\equiv\sin{\beta_0};\quad\cz\equiv\cos{\beta_0};\quad\etaz=\sz\cz=\tfrac{1}{2\sqrt{2}}\quad{\rm and}\quad\rhoz=\sz^2.\label{zeroConstants}
\eeq
\noindent We illustrate \eqs{z1u}{justification} in figure~\ref{fig1}, which shows the leading-order UT (``LO-UT'') whose angles and sides are used as input to our texture, \eq{tex1}, which ensures that an approximately congruent UT results after diagonalisation.
 The apex of the standard UT is simply given by  \cite{pdg,burasetal}:
\begin{align}
\rhob+i\etab\simeq\zz.
\label{intermsofrhoeta}
\end{align}

Our working so far has been in the small-angle limit, i.e.~ignoring cosines of the quark mixing angles, so that we cannot assume a relative accuracy better than $1-\cos(\theta_{C})\sim\lam^2/2\sim2.5\%$. Thus, we have not yet needed to distinguish between the two pairs of similar parameters, $(\rho,\eta)$ \cite{wolfenstein} and $(\rhob,\etab)$ \cite{burasetal}, which differ by just this fraction \cite{pdg}. For the numerical results below, following the PDG \cite{pdg}, we choose to work with $(\rhob,\etab)$. The full next-to leading order (NLO) algebraic solutions of the texture for the mixings and mass ratios are given in the Appendix.

\section{Numerical fit to the data}
For the fit to the data, a full diagonalisation is performed at high numerical precision. The CKM observables are taken from the latest PDG summary \cite{pdg}. Significant recent progress in quark mass determinations using lattice QCD \cite{flag} has led to dramatic reductions in the uncertainties on the quark mass ratios (which we take as experimental inputs\footnote{Lattice quark mass determinations are calibrated to experimentally-determined meson masses.}). We follow the approach of \cite{bb}, updating the inputs with the latest lattice values compiled in the review on quark masses \cite{pdg}. These inputs are quoted in the second column of table \ref{table:fits} renormalised to the scale $\mu=m_t=173.1$ GeV in the $\overline{\rm MS}$ scheme. A \chisq\ fit to the data is performed, varying the five free parameters of the MMs to find their preferred values.
\begin{table}[h]
\centering
\def\arraystretch{1.3} 
{\setlength{\tabcolsep}{0.5em}
\begin{tabular}{ c | c  c  c  c }
  		Observable 		& Experimental 		& Input Renorma-			& Fitted Value   		& Pull   		\\
  				 		& Input at $m_t$		& lised to $10^4$ TeV 			&  at $10^4$ TeV   & 		   	\\
\hline
$|m_u/m_c|\,(\times 10^{3})$ 	& \ruExp		 		& Unchanged	& $\ruFit~~~~~~$ 			& $~~\ruPull$   	\\
$|m_d/m_s|\,(\times 10^{2})$	& \rdExp		 		& \ditto		& $\rdFit~~~~~~$ 			& $~~\rdPull$ 	\\
$m_c/m_t\,(\times 10^{3})$ 	& \rcExp		  		& \rcRenorm	& $\rcFit~~~~~~$ 			& $\rcPull$   	\\
$m_s/m_b\,(\times 10^{2})$ 	& \rsExp		  		& \rsRenorm	& $~\,\rsFit~~~~~~$ 		& $~~\rsPull$  		\\
$\lam$ 					& \lamExp			 	& Unchanged	& $~~~~\lamFit~~~~~~$ 	& $\lamPull$  		\\
$A$ 						& \AExp~~			& \ARenorm	& $~\AFit~~~~~~$ 		& $\APull$			\\
$\rhob$ 					& \rbExp 				& Unchanged	& $~~~\rbFit~~~~~~$ 		& $\rbPull$  	\\
$\etab$ 					& \etabExp~~~ 			& \ditto		& $~~~\etabFit~~~~~~$ 	& $\etabPull$  		\\
\hdashline
Total $\chi^2/dof$ 			& $-$ 				& $-$		& $~~~\chisqFit/\nDof~~~~~~$ 	& $~-$   		\\
\hdashline
$\alpha\, (^{\circ})$ 			& \alphaExp		 	& Unchanged	& $~~\,~~91.30\pm0.02$ 		&   $-0.21$ 	\\
$\beta\, (^{\circ})$ 			& $22.6\pm0.4$ 		& \ditto		& $~\,22.3\pm0.1$ 		& $-0.75$   		\\
$\gamma\, (^{\circ})$ 		& \gammaExp			& \ditto		& $~\,66.4\pm0.1$ 		& $~~\,0.54$   		\\
$\,\beta_s\, (^{\circ})$ 		& $1.15\pm0.46$		& \ditto		& $~~~1.07\pm0.01$ 	& $\,-0.17$   		\\
\end{tabular}
}
\caption[]{Comparison of observed quark mass ratios and mixing observables at $m_t$ and renormalised to $10^4$ TeV, with those obtained by the fit to the textures, \eq{tex1}. The predictions shown for the UT angles in the penultimate column apply also at the weak scale, since they have negligible evolution with renormalisation scale. $\beta_s$ is determined from $CP$-violation in $B^0_s$ decays \cite{pdg}. Pulls are calculated as the difference between the central fitted value and the renormalised input values, as a fraction of the renormalised experimental errors. The values of the fitted parameters are: $\aa=0.854\pm0.013$, $\bb=0.462\pm0.001$, $\dd=-0.040\pm0.006$, $\du=0.344\pm0.003$ and $\xx=0.22646\pm0.00034$.}
\label{table:fits}
\end{table}

The fit to the data renormalised at the weak scale ($\mu=m_t$, not shown), is excluded by its $\chisq/dof$ ($\simeq 100/3$). However, the lack of observations of new physics at around this scale makes it unlikely anyway that a MM texture applies at this scale. Several of the fitted observables ($A$, $m_c/m_t$ and $m_s/m_b$) evolve with renormalisation scale, and the rate and direction of this evolution in the SM are well-known \cite{smRenorm1,smRenorm2,smRenorm3,ckmRenorm1,ckmRenorm2}. It is thus possible that our texture might more naturally apply at some higher energy scale, so we have scanned the SM evolution of the relevant observables up to the GUT scale, to investigate this possibility. The best fit is at a scale $\mu\sim 10^4$ TeV with a $\chisq/dof$\footnote{While the scale $\mu$ is not a parameter of the texture, we include it in this count of fit parameters.} $\simeq1.0/2$ and a width of $\sim$ a factor 3 on either side, giving the range \hbox{$\mu\sim(0.3\rightarrow3)\times10^4$ TeV.}

The full fit results for the preferred solution are shown in table \ref{table:fits} where the second column shows the input data, the third gives the same inputs renormalised to the best fit scale, $\mu\sim 10^4$ TeV, the fourth gives the fitted values of the observables, while the fifth gives the individual pulls compared with the input observables at the best fit scale. The $\chi^2$ is calculated summing over those parameters above the dashed line for all observables renormalised to the best fit scale.  The best fit parameter values and their errors are given in the table caption. Observables shown below the total $\chi^2$ line are not independent, and are not used in the fit, but are given for completeness as scale-independent predictions. Their values may be tested more precisely than at present by future measurements.

As is usual, the signs of the masses are unphysical, so that different signs of the fitted mass ratios must be allowed
for in the fit. Only two of the possible sign permutations have viable \chisq\ values, i.e.~those with either sign of $m_u/m_c$, with $m_d/m_s$ negative and the other mass ratios positive. These two solutions are mutually unresolved, being almost degenerate in $\chi^2$, and in the values of all fit parameters except $\du$, which controls $m_u/m_c$. To avoid repetition, we have reported only the fit for the smaller value of $\du$ (with $m_u/m_c$ negative). The fits for other mass-sign permutations are strongly disfavoured by their $\chisq$ values.

\section{Constraints}
Naturally, with five parameters and eight observables, there are, in principle, three constraints among the observables. At LO in $\xx$, they are independent of any of the parameters, whereas at higher orders, they become parameter-dependent. From \eq{secondRatio} and \eqs{zAns0}{intermsofrhoeta} we combine to give the LO constraints:
\bal
\frac{\mc}{\mt}\frac{\mb}{\ms}=&\,\frac{\zul^2}{\zdl^2}=\tan^2{\pibyeight}=0.172,~{\rm c.f.}~0.177\pm0.002\,(exp)=0.176\,(fit);\label{doubleRatio}\\
\etab=\etaz&\equiv\tfrac{1}{2\sqrt{2}}=0.354,~{\rm c.f.}~0.352\pm0.007\,(exp)=0.348\,(fit);\label{etaPred}\\
\rhob+\etab&=\rhoz+\etaz\equiv\half,~{\rm c.f.}~0.511\pm0.012\,(exp)=0.500\,(fit),\label{rhoPlusEta}
\end{align}
where \eq{doubleRatio} is a scale-dependent prediction and \eqs{etaPred}{rhoPlusEta} are scale-independent. The numerical results, given for comparison above, are from table \ref{table:fits} and are given at the best fit scale, $\mu\sim10^4$ TeV. The analytic NLO formulae are given in the Appendix, \eqs{eqRho}{m2Overm3}, and can be combined to provide a better match to the fitted values quoted above. Finally, we note the suggestive similarity between \eq{doubleRatio} and the phenomenologically-successful prediction for the neutrino mass ratio: $\frac{m_2}{m_3}=\tan^2\!{\pibyeight}$ in \cite{krishnan1}.

\section{Symmetries of the mass matrices}
The texture, eqs.~(\ref{tex1}) and (\ref{zrat}), and its fit, table \ref{table:fits}, are the main results of this paper. Having established that a viable fit to the data is possible for this texture, we identify the two invariance properties of the MMs which realise the complex constant, $\zu/\zd$, defined in \eq{zrat}.

The ratio $\zu/\zd$ involves a parameter from each MM, so that a regularity therein cannot be reflected in either matrix alone. The UT is significant in providing a diagrammatic representation of a relationship \emph{between} the MMs and we will find it particularly useful in illustrating the symmetries inherent in our MM-pair. We first recall that a $CP$ transformation is effected by a simple complex conjugation of the MMs: here, simply $\zz\rightarrow \zz^*,~\zb\rightarrow \zb^*$. In consequence, the Jarlskog invariant, $\JCP$, flips sign, the UT is reflected in the real axis, and other observables are preserved. Naturally, a very poor fit to the data results, due to the resulting wrong sign of $CP$-violating observables. Less obviously, with our textures, under either $\zz\rightarrow-\zz$ or $\zb\rightarrow-\zb$ (but not both), we also obtain a sign-flip of $\JCP$, and effectively (see below) a reflection of the UT in the real axis. Thus the simultaneous application of both $\zz\rightarrow-\zz$ and a $CP$ transformation cancel each other out, making the {\em combined} transformation a symmetry of our MM-pair. Analogously, the simultaneous application of $\zb\rightarrow-\zb$ together with $CP$ likewise leaves the observables invariant, an alternative manifestation of the same  symmetry.

\begin{figure}[t]
\begin{center}
\includegraphics[scale=1.10]{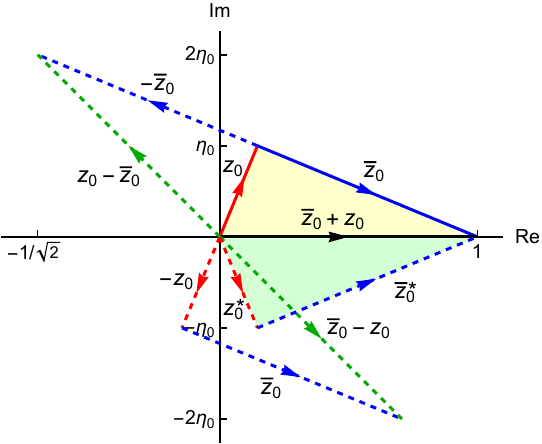}
\caption{\label{fig:symm1}
The leading-order UT, shaded yellow, and its $CP$ transform (complex conjugate) in green. The two unshaded triangles, formed using $(-\zz,\zb)$ and $(\zz,-\zb)$ respectively, are both exactly congruent to the green one, each under a simple rotation, but only if the phase-difference between $\zz$ and $\zb$ is $\pibytwo$. Thus, in this case, they are equivalent to the $CP$ transformation, since rotations of the UT are unobservable. This illustrates the invariance under the combined transformation of sign flip (of $\zz$ or $\zb$) and $CP$.}
\end{center}
\end{figure}

The situation is illustrated in figure~\ref{fig:symm1},
where the LO-UT is again shaded yellow, with its complex-conjugate (i.e.~its $CP$-transform) green. We recall that simultaneous phase changes of the off-diagonal elements of the MMs are unobservable and correspond to rotations of the unitarity triangle in the complex plane \cite{JarlsRot1,JarlsRot2,JarlsRot3,JarlsRot4,JarlsRot5}. In the figure, we can see that a flip in the sign of $\zz$ alone results in a new LO-UT (unshaded) which is congruent to the $CP$-transformed triangle, being  rotatable into it. The example with the triangle formed using $\zz$ and $-\zb$ behaves similarly. Figure~\ref{fig:symm1} thus makes manifest the symmetry under simultaneous $CP$ transformation and single sign flip of either $\zq$ (but not both).

We emphasise that this symmetry is a direct result of the $\pibytwo$ phase difference between $\zz$ and $\zb$. It is clear from figure~\ref{fig:symm1} that if $\zz$ and $\zb$ had any other phase difference, then forming a triangle using $-\zz$ and $\zb$ or ($\zz$ and $-\zb$) would result in a triangle which was not congruent to the $CP$-transformed one, indicating that then, the symmetry is absent. One could reverse the argument and say that requiring such an invariance in the MMs is sufficient to ensure that the two $\zq$ have a relative phase of $\pibytwo$. The situation here is reminiscent of that in many models of lepton mixing wherein invariance under a combined transformation involving $CP$ and a second involution (e.g.~$\mu-\tau$ exchange in \cite{mutauSymm}) results in a leptonic $CP$-phase close to $\pibytwo$. 

The above considerations extend to the case of the UT resulting from full diagonalisation, despite its being slightly perturbed relative to the LO-UT illustrated. This is verified using Jarlskog's $CP$-violating determinant \cite{JarlskogDet1,JarlskogDet2}, which, like the UT, {\em links} both MMs:
\beq
\DD\equiv Det[M^0_u,M^0_d]=-2iF_uF_d\Jcp,\\
\label{Jdet}
\eeq
where $F_q\simeq(m^q_1-m^q_2)(m^q_2-m^q_3)(m^q_3-m^q_1)/(m_3^q)^3$. The determinant $\DD$ is readily shown to flip sign under either of the two $\zq$ (i.e.~$\zz$ or $\zb$) sign reversals separately (given a $\zq$ phase difference of $\pibytwo$), and of course under $CP$ transformation. Interestingly, this latter check only works in the case of a texture in which the 13 and 31 elements of the MMs are zero, linking this invariance also to the texture zeroes. Thus, it seems that requiring invariance of the observables under the compound transformation also helps to constrain the form of the texture, beyond just the phase relationship of the $\zq$.

\begin{figure}[t]
\begin{center}
\includegraphics[scale=0.86]{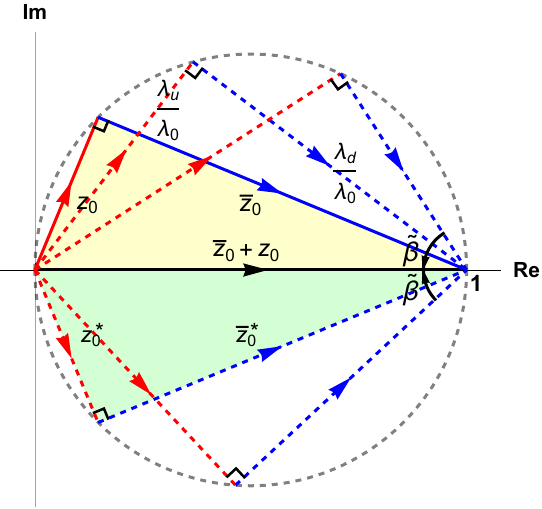}
\caption{\label{fig3}
Leading-order UT, for our texture-pair (yellow, $\bt=\pibyeight$), and (unshaded) three possible generalisations which represent a continuum with variable opening angle $\bt$ in the range $-\pibytwo<\bt<\pibytwo$ (negative values flip the sign of $\JCP$). The $CP$-conjugate LO-UT ($\bt=-\pibyeight$), in green, is reachable by the particular $\bt$ transformation, \eq{rotPibytwo}, i.e.~$\bt\ra \bt-\pibyfour$, but only because $\bt=\bz=\pibyeight$. The figure is drawn in the PDG phase convention, although, since $\bt$ is a relative phase, the figure can be rotated between phase conventions. Note that $|\zb+\zz|=1$ for all $\bt$ and for all phase conventions.}
\end{center}
\end{figure}

We turn now to consider how the ratio $\zul/\zdl$ may be constrained to the value $\tan{\pibyeight}$ by means of another invariance of the MMs. It is useful first to consider the more general situation where this is not the case (but the $\zq$ still retain their $\pibytwo$ phase difference). Again the UT and Jarkskog's determinant are instrumental to the argument. Figure~\ref{fig3} illustrates the LO-UT for this relaxed texture, and we see that:
\beq
\left|\frac{\zu}{\zd}\right|=\left|\frac{\zz^*}{\zb^*}\right|=\left|\frac{\zz}{\zb}\right|=\tan{\bt},
\label{tanBtilde}
\eeq
where $\bt$ is the $CP$-violating opening angle between $\zb$ and $\zz+\zb$ in the complex plane (such that we still have the UT angle $\beta\simeq \bt$ after diagonalisation).

The Jarlskog determinant, \eq{Jdet}, is of course $CP$-violating and is a function of $\bt$. We find, for the generalised texture, the exact expression:
\begin{equation}
\DD=-i\xx^{10} \bb^2 \aa^2 \sin^3\!{(2\bt)}f_{\DD}(b,\du,\dd,\xx,\cos{2\bt})
\label{DD}
\end{equation}
($-\pibytwo<\bt<\pibytwo$), where $f_{\DD}(b,\du,\dd,\xx,\cos{2\bt})$ is a polynomial in $\cos(2\bt)$. We consider now the following (observable) rotation of $\zb$ in the complex plane:
\beq
2\bt\rightarrow2\bt-\pibytwo.
\label{rotPibytwo}
\eeq
Under the transformation, \eq{rotPibytwo}, we have that $\sin{2\bt}\rightarrow -\cos{2\bt}$ and $\cos{2\bt}\rightarrow \sin{2\bt}$, which obviously changes the value of $\DD$, \eq{DD}, in general. However, in the particular case that $\bt=\bz\equiv\pibyeight$, the only effect is to flip the sign of $\sin{2\bt}$ (and hence that of $\DD$), effectively a $CP$ transformation. This invariance can also be verified by inspecting figure~\ref{fig3}. Symmetry can be restored again by performing a further explicit $CP$ transformation (exactly analogous to the invariance under the compound $\zq$ sign flip and $CP$ transformation discussed earlier).
\emph{Thus the symmetry of the texture which ensures} $\bt=\bz\equiv\pibyeight$ \emph{is under the combined transformation, \eq{rotPibytwo} and $CP$}.

\section{Discussion}
In summary, we have introduced a novel texture for the two normalised quark mass matrices, having five free parameters in total. We use a small expansion parameter similar to Wolfenstein's, but, motivated by the UT, split here into two distinct complex components, $\zu$ and $\zd$, whose complex sum has magnitude $\lam+{\cal O}(\lam^3)$. These complex components have a fixed ratio $\zu/\zd=-i\tan{\pibyeight}$, which yields a UT consistent with experiment. In this ``geometric'' ansatz, the distinctiveness of the mass hierarchies for the two quark sectors, that of the ``up'' quarks being more extreme than that of the ``down'' quarks, is directly linked to the different magnitudes of the two sloped sides of the UT (i.e.~$|V_{ub}|<|V_{td}|$) via the relation $|\zu/\zd|\simeq|V_{ub}|/|V_{td}|\simeq\tan\beta<1$, which controls both. We obtain a good fit to the data renormalised to a scale in the range $(0.3-3)\times10^4$ TeV.

We do not propose an extension of the SM with additional fields and quantum numbers to underpin the proposed texture, restricting our analysis to the phenomenology of the MMs themselves. We have instead elucidated two distinct new symmetries, manifested by the MMs, but absent from the SM, each involving an elementary transformation of the $\zq$ alone, with symmetry restored in each case by a $CP$ transformation. We have shown that these symmetries are necessary and sufficient to constrain the UT angles to the experimentally-determined values. Whether these symmetries occur in isolation, or are a manifestation of some higher symmetry is, for the time being, unknown.

The weak isospin reflection symmetry among the coefficients of the MMs is needed to reduce the number of parameters to make the ansatz more predictive, and is consistent with the data. The fact that two distinct parameters are needed in the smallest `11' entry of the MMs is the least appealing aspect of the texture, but it is perhaps not entirely surprising, since radiative corrections due to the gauge interactions might dominate the Yukawa contribution in this very small element and cannot be expected to be isospin-symmetric in the normalised mass matrices. Notwithstanding this feature, the proposed textures fit the experimental results and reduce the number of parameters needed to describe the data in the quark flavour sector from ten to seven (including the normalisations of the MMs).

A potentially useful and novel practical feature of the texture-pair is that among its input parameters and its fixed (by symmetry) constants, four of these seven quantities are directly and transparently seen to be equal in leading approximation to recognizable mixing observables, namely, $\xx\simeq\lam$, $\aa\simeq A$, $\rhoz\simeq\rho$, $\etaz\simeq\eta$. The two identified compound symmetries are associated with the phenomenologically-successful UT relations $\alpha\simeq\pibytwo$ and $\beta\simeq\pibyeight$ respectively, and we have made precise numerical predictions for the angles of the UT (given in table \ref{table:fits}), which can be tested by future experimental measurements.

\section*{Acknowledgements}
It is a pleasure to thank Tim Gershon, Rama Krishnan, Chris Sachrajda and Jack Shergold for useful discussions and/or helpful comments on the manuscript. PFH acknowledges support from the UK Science and Technology Facilities Council (STFC) on the STFC consolidated grant ST/W000571/1.

\appendix
\section{Analytic NLO solutions}
For completeness, we provide here the algebraic NLO solutions of the texture:
\bal
\lam&=\xx\left(1+\flam\xxSq\,\right)+{\cal O}(\xxFive)\label{lamNLO}\\
A&=\aa\left\{1+\left[\tfrac{1}{4}(3\bb-2\rhoz)-2\flam\right]\xxSq\,\right\}+{\cal O}(\xxFour)\label{AprimeNLO}\\
\rhob&=\rhoz\left(1+\cz\frho\xxSq\,\right)+{\cal O}(\xxFour)\label{eqRho}\\
\etab&=\etaz\left\{1+\left[\sz\frho + \tfrac{1}{2}(1 - 5 \bb)\right]\xxSq\,\right\}+{\cal O}(\xxFour),\label{etaNLO}
\end{align}
where $\flam\equiv\tfrac{1}{4}(3\fA+4\etaz\fc-5)\simeq-0.11$, $\fA\equiv\tfrac{1}{\bb}\!\left[\aaSq+\half(\dd+\du)\right]$, 
$\fc\equiv\tfrac{1}{\bb}(\dd-\du)$, and $\frho\equiv\sz(1+\fc)-\tfrac{1}{2\sz}\!(\fA + \fc - \tfrac{7}{2} \bb)\simeq0.76$.
For the mass ratios, we find:
\bal
\frac{m^q_1}{m^q_2}&=-\zql^2(1-r_q)\left\{1+\left[r_A\tfrac{(2-r_q)}{(1-r_q)}-2\right]\zql^2\right\}+{\cal O}(\zql^6)\label{m1Overm2}\\
\frac{m^q_2}{m^q_3}&=\bb\zql^2\left[1+(1-r_A)\zql^2\right]+{\cal O}(\zql^6),\label{m2Overm3}
\end{align}
where $r_q=\tfrac{\dq}{\bb}$ and $r_A=\tfrac{\aa^2}{\bb}$.
The NLO corrections in \eqs{lamNLO}{etaNLO} as fractions of the LO terms are respectively: $-5.8\times10^{-3}$, $+2.6\%$, $+3.6\%$ and $-1.8\%$, while the NLO corrections in \eqs{m1Overm2}{m2Overm3} to the mass ratios $m_c/m_t$, $m_s/m_b$, $m_u/m_c$, $m_d/m_s$ as a fraction of the LO terms are respectively $-4.3\times10^{-3}$, $-2.5\%$, $+4.3\%$, and $+4.5\%$ (all calculated using the fitted values reported in table 1). In all cases, the results are compatible with the full numerical results reported in table 1, given the indicated degree of approximation.

\end{document}